\documentclass[twocolumn]{aastex62}
\journalinfo{\@Draft}
\usepackage{natbib}
\usepackage{xcolor}
\usepackage{color}
\usepackage{comment}
\usepackage{amsmath}
\definecolor{color}{RGB}{25,25,112}
\definecolor{negro}{RGB}{0,0,0}
\definecolor{colorurl}{RGB}{25,25,112}
\renewcommand\tablename{Table}
\usepackage{hyperref}
\hypersetup{
  colorlinks,
  citecolor=color,
  linkcolor=color,
  urlcolor=colorurl}
\shorttitle{GRB 180620A: EVIDENCE FOR LATE-TIME ENERGY INJECTION}

\begin{document}

\title{GRB 180620A: EVIDENCE FOR LATE-TIME ENERGY INJECTION}

\author{Becerra,~R.~L.}
\affiliation{Instituto de Astronom{\'\i}a, Universidad Nacional Aut\'onoma de M\'exico, Apartado Postal 70-264, 04510 M\'exico, CDMX, M\'exico;}

\author{De~Colle,~F.}
\affiliation{Instituto de Ciencias Nucleares, Universidad Nacional Aut\'onoma de M\'exico, Apartado Postal 70-264, 04510 M\'exico, CDMX, M\'exico;}

\author{Watson,~A.~M.}
\affiliation{Instituto de Astronom{\'\i}a, Universidad Nacional Aut\'onoma de M\'exico, Apartado Postal 70-264, 04510 M\'exico, CDMX, M\'exico;}

\author{Fraija,~N.}
\affiliation{Instituto de Astronom{\'\i}a, Universidad Nacional Aut\'onoma de M\'exico, Apartado Postal 70-264, 04510 M\'exico, CDMX, M\'exico;}

\author{Butler,~N.~R.}
\affiliation{School of Earth and Space Exploration, Arizona State University, Tempe, AZ 85287, USA;}

\author{Lee,~W.~H.}
\affiliation{Instituto de Astronom{\'\i}a, Universidad Nacional Aut\'onoma de M\'exico, Apartado Postal 70-264, 04510 M\'exico, CDMX, M\'exico;}

\author{Rom\'an-Z\'u\~niga,~C.~G.}
\affiliation{Instituto de Astronom{\'\i}a, Universidad Nacional Aut\'onoma de M\'exico, Unidad Acad\'emica en Ensenada, 22860 Ensenada, BC, Mexico;}

\author{Bloom,~J.~S.}
\affiliation{Department of Astronomy, University of California, Berkeley, CA 94720-3411, USA;}

\author{Gonz\'alez,~J.~J.}
\affiliation{Instituto de Astronom{\'\i}a, Universidad Nacional Aut\'onoma de M\'exico, Apartado Postal 70-264, 04510 M\'exico, CDMX, M\'exico;}

\author{Kutyrev,~A.~S.}
\affiliation{Space Telescope Science Institute, 3700 San Martin Drive, Baltimore, MD 21218, USA;}

\author{Prochaska,~J.~X.}
\affiliation{Department of Astronomy and Astrophysics, UCO/Lick Observatory, University of California, 1156 High Street, Santa Cruz, CA 95064, USA;}

\author{Ramirez-Ruiz,~E.}
\affiliation{Department of Astronomy and Astrophysics, UCO/Lick Observatory, University of California, 1156 High Street, Santa Cruz, CA 95064, USA;}

\author{Richer,~M.~G.}
\affiliation{Instituto de Astronom{\'\i}a, Universidad Nacional Aut\'onoma de M\'exico, Unidad Acad\'emica en Ensenada, 22860 Ensenada, BC, Mexico;}

\author{Troja,~E.}
\affiliation{Astrophysics Science Division, NASA Goddard Space Flight Center, 8800 Greenbelt Road, Greenbelt, MD 20771, USA;}
\affiliation{Department of Astronomy, University of Maryland, College Park, MD 20742-4111, USA}

\begin{abstract}

The early optical emission of gamma-ray bursts gives an opportunity to understand the central engine and first stages of these events. About 30\% of GRBs present flares whose origin is still a subject of discussion.
We present optical photometry of GRB 180620A with the COATLI telescope and RATIR instrument. COATLI started to observe from the end of prompt emission at $T+39.3$~s and RATIR from $T+121.4$~s. We supplement the optical data with the X-ray light curve from \emph{Swift}/XRT.
We observe an optical flare from $T+110$ to $T+550$~s, with a temporal index decay $\alpha_\mathrm{O,decay}=1.32\pm 0.01$, and a $\Delta t/t=1.63$, which we interpret as the signature of a reverse shock component. 
After the initial normal decay the light curves show a long plateau from $T+500$ to $T+7800$~s both in X-rays and the optical before decaying again after an achromatic jet break at $T+7800$~s. Fluctuations are seen during the plateau phase in the optical. 
Adding to the complexity of GRB afterglows, the plateau phase (typically associated with the coasting phase of the jet) is seen in this object after the ``normal'' decay phase (emitted during the deceleration phase of the jet) and the jet break phase occurs directly after the plateau.
We suggest that this sequence of events can be explained by a rapid deceleration of the jet with $t_d\lesssim 40$ s due to the high density of the environment ($\approx 100$ cm$^{-3}$) followed by reactivation of the central engine which causes the flare and powers the plateau phase.
\end{abstract}

\begin{center}
\end{center}

\section{Introduction}
\label{sec:introduction}

 Gamma-ray bursts (GRBs) are the brightest events observed in the universe. The duration of GRBs is measured by the $T_{90}$ parameter which is defined as the time interval during which 90\% of the total observed counts are detected. We observe two populations of GRBs with distinctive distributions of $T_{90}$. Long GRBs (whose duration is typically $T_{90}>2$ s) are the result of the collapse of massive stars \citep[e.g.,][]{1993ApJ...3505..273W,mw99} whereas  short GRBs (whose duration is typically $T_{90}<2$ s) are the result of a coalescence of two compact objects \citep[e.g.,][]{ls76,bp86,bp91,2007NJPh....9...17L}.

The fireball model, the standard theory which describes most of the features of GRBs, distinguishes two main stages. First, the prompt emission, which is caused by the dissipation of kinetic energy in internal shocks \citep{1994ApJ...430L..93R,1997ApJ...490...92K} and/or by photospheric emission \citep{1994MNRAS.270..480T,2000ApJ...529..146E}. Second, the afterglow phase, generated by the deceleration of the jet due to its interaction with the circumburst medium.  This deceleration leads to the formation of reverse and  forward shocks, across which the jet and the ambient medium (in the frame of the shocked material) 
are decelerated and heated \citep[e.g.,][]{1999PhR...314..575P}.

In most cases, the evolution of the afterglow may be summarized as follows: a steep decay phase related with the end  of the prompt emission phase, a shallow decay phase (or plateau), a ``normal'' decay phase, a jet-break steepening, and flares in some but not all bursts \citep[see, e.g.,][]{2006ApJ...642..354Z,2009MNRAS.397.1177E,2006ApJ...642..389N}

About 20\% of long GRBs present an optical rebrightening in the afterglow \citep[e.g.,][]{2012ApJ...758...27L}. They are commonly described by a FRED (fast-rise and exponential-decay) behaviour. Their spectral and temporal indices seem to indicate that they share a common origin with the
prompt emission, in good agreement with that 
observed in X-rays \citep{2005ApJ...631..429I}. This behaviour suggests a reactivation of the central engine. Nevertheless,  anisotropic or ambient density fluctuations, as well as a reverse shock component are also able to produce a rebrightening in the light curve. These different models predict different values for the duration of the flare $dt$ over the peak time $t$, and we can use this criterion to distinguish the process involved \citep{2005ApJ...631..429I}. A detailed temporal study of these flares can improve the understanding of early stages of GRBs.

Catching early optical emission from GRBs is not an easy task. The short duration of these events, combined with telemetry delays and the response time of ground telescopes and satellites as \emph{Swift}/UVOT, present a challenge. For these reasons, our understanding of the earlier phases of GRBs continues to be incomplete compared to our understanding of later afterglows (occurring minutes after the trigger) for which we have a sample of hundreds of observations \citep[e.g., see][]{2007ApJ...655..391K,2015PhR...561....1K, 2017ApJ...848...15F}.
The main contributions of our current paper are to add data to the sample of early optical emission photometry of GRBs and to provide a detailed physical interpretation. 

GRB 180620A was one of the brightest GRBs in the last few years. It presented a bright optical counterpart and it exhibited a temporal evolution never seen before for a GRB. It was well observed by many collaborations. A total of 16 GCN Circulars related to GRB 180620A were published between June 20 and June 25, 2018. In this paper we present new optical photometry of the bright GRB 180620A with the COATLI telescope in the $w$-filter and RATIR in the $r$- and $i$-filters from early to late stages. The GRB 180620A event posed an excellent opportunity for improving our understanding of early emission and flares of GRBs.

The paper is organized as follows. In \S\ref{sec:observations}, we present the observations with COATLI and RATIR. We present our results in \S\ref{sec:analysis} and a physical interpretation in \S\ref{sec:interpretation}. Finally, we discuss the scenario that we propose to interpret our data in \S\ref{sec:discussion} and summarize our conclusions in \S\ref{sec:summary}. 

\section{Observations}
\label{sec:observations}

\subsection{Neil Gehrels Swift Observatory}
\label{sec:swift}

The {\itshape Swift}/BAT instrument triggered on GRB 180620A at $T =$ 2018 June 20 08:34:58 UTC \citep{22798}. The {\itshape Swift}/BAT on-board  location was 18:39:32 +23:15:55 J2000, with an uncertainty of 3 arcmin (radius, 90\% containment, including systematic uncertainty). The {\itshape Swift}/BAT light curve showed several overlapping peaks with a total duration of about 25~s.
There were two main peaks at $T + 5$~s and $T + 12$~s, with the later peak being stronger \citep{22798}. The duration of GRB 180620A  (15-350 keV) was $T_{90}=23.16\pm$ 4.82 seconds \citep{22815}, making it a long GRB.

The {\itshape Swift}/XRT instrument started observing the field at $T+267.1$~s and found an uncatalogued X-rays source at 18:39:34.98 +23:14:37.4 J2000 within 3.6 arcsec (radius, 90\% containment). This location is 87 arcsec from the BAT on board position and within the BAT error circle \citep{22798}. The fluence in the 15-150 keV band was 5.8$\pm$0.2$\times10^{-6}$ erg cm$^{-2}$. 

The {\itshape Swift}/UVOT took a finding chart exposure of nominal 150 seconds with the white
filter starting at $T+270$~s \citep{22798}, and began settled observations of the field of GRB 180620A at $T+270$~s \citep{22805}. The detection by the {\itshape Swift}/UVOT instrument implied a redshift limit of 1.2.

\subsection{COATLI Observations}
\label{sec:observationscoatli}

COATLI\footnote{\url{http://coatli.astroscu.unam.mx/}} is a robotic 50-cm telescope at the Observatorio Astron\'omico Nacional on Sierra de San Pedro M\'artir in Baja California, Mexico \citep{2016SPIE.9908E..5OW}. COATLI is connected to the GCN/TAN alert system and received the {\itshape Swift}/BAT alert packet for GRB 180620A at 08:35:13.2 UTC ($T + 15.2$~s). It immediately slewed to the burst and began observing, with the first exposure starting at 08:35:37.3 UTC which corresponds to $T+ 39.3$~s and observed the field of GRB 180620A up to 2.7 hours after the trigger, obtaining a total of 6250 seconds of exposure in the $w -$filter \citep{22806}. The delay between receiving the alert and the first observation was $21.1$~s; COATLI can often respond more quickly, but in this case its German equatorial mount had to perform a meridian flip which creates an additional delay.

Our reduction pipeline performs bias subtraction, dark subtraction, flat-field correction and cosmic-ray cleaning with the \emph{cosmicrays} task in IRAF \citep{1986SPIE..627..733T}. To coadd the later images, we measured the offsets between the brightest star of the field and then used the  \emph{imcombine} routine of IRAF \citep{1986SPIE..627..733T}.
We performed astrometric calibration of our images using the \url{astrometry.net} software \citep{2010AJ....139.1782L}. 

For times $T < 550$~s, COATLI observations are 5-second exposures in the clear $w$ filter. The read time for the CCD is about 4 seconds, so the cadence was typically about 9 seconds. The telescope dithered, taking ten images at one dither position before moving to the next dither position. From $T + 550$~s to $T + 1800$~s, we combine sets of 10 exposures taken over about 86 seconds. From $T + 9000$~s to the end of the first night we combine sets of 10 exposures taken over about 300 seconds to improve the signal-to-noise ratio.

We performed aperture photometry using Sextractor \citep{1996A&AS..117..393B} with an aperture of 3.5 arcsec diameter. Our measurements were calibrated against Pan-STARRS DR1. Table~\ref{tab:datoscoatli} summarizes our COATLI photometry. For each image it lists the start and end times of the observation, $t_i$ and $t_f$, relative to the trigger time $T$,  the total exposure time, $t_{exp}$, the AB magnitude $w$ (not corrected for Galactic extinction), and the 1$\sigma$ total uncertainty in the magnitude (including both statistical and systematic contributions).

\begin{figure*}
\centering
 \includegraphics[width=0.85\textwidth]{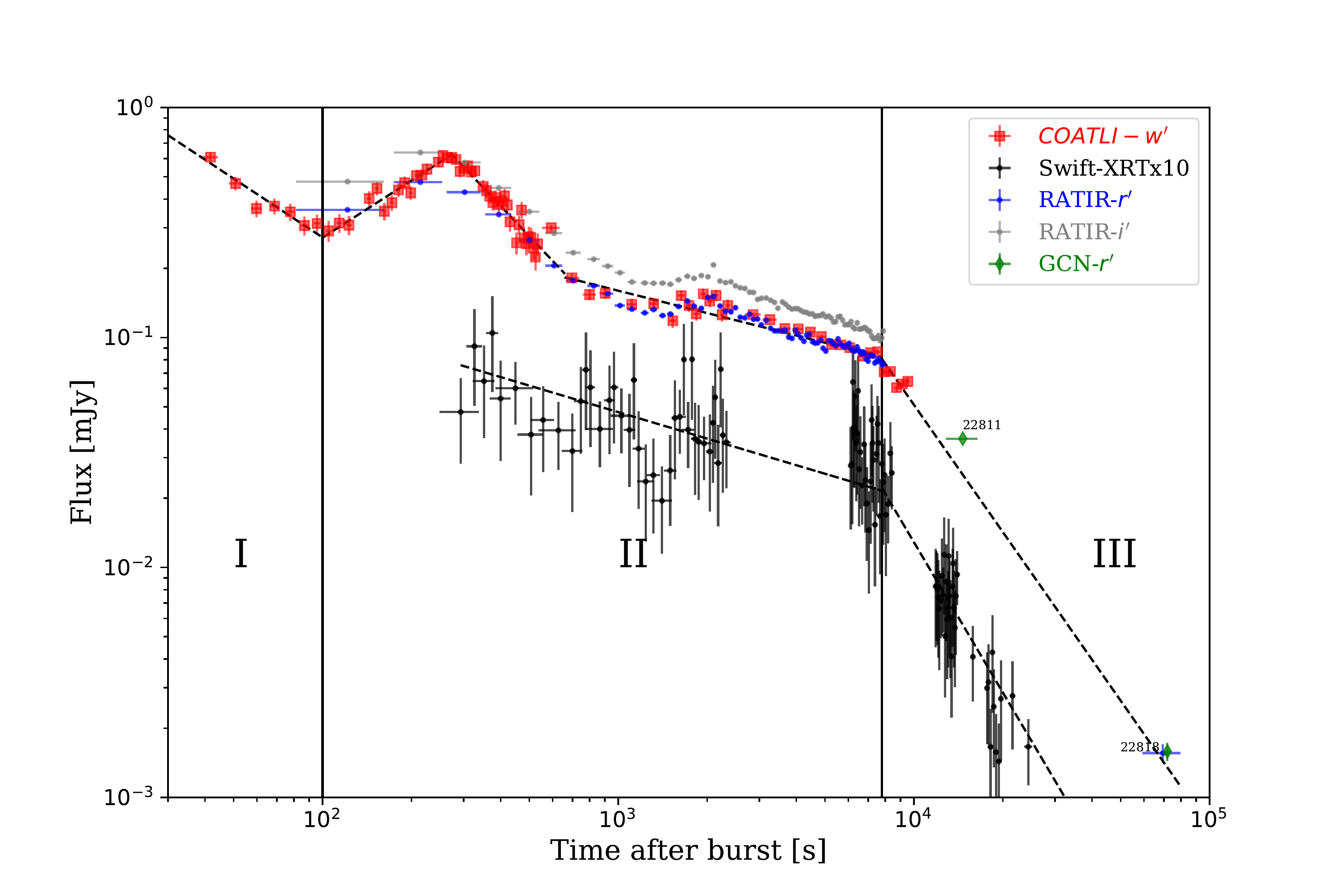}
 \caption{Light curves and broken power-law fits of GRB 180620A in $w$ from COATLI (red squares), $r$ and $i$ from RATIR (blue and grey points respectively) and X-rays from {\itshape Swift}/XRT (black points) at 1~keV. We added photometry of GCNs from \cite{22811} and \cite{22818} (green points). We divided the light curve in three different stages, corresponding to I) normal decay phase, II) plateau phase, and III) steep decay phase.}
 \label{fig:lightcurve}
\end{figure*}

\subsection{RATIR Observations}
\label{sec:observationsratir}

RATIR\footnote{\url{http://ratir.astroscu.unam.mx/}} is a four-channel simultaneous optical and near-infrared imager mounted on the 1.5 meter Harold L. Johnson Telescope at the Observatorio Astron\'omico Nacional in Sierra San Pedro M\'artir in Baja California, Mexico. RATIR responds autonomously to GRB triggers from the Swift satellite and obtains simultaneous photometry in $riZJ$ or $riYH$ \citep{2012SPIE.8446E..10B,2012SPIE.8444E..5LW,2015MNRAS.449.2919L}. For the observations of GRB 180620A, the $ZYJH$ channels were not available.

The reduction pipeline performs bias subtraction and flat-field correction, followed by astrometric calibration using the \url{astrometry.net} software \citep{2010AJ....139.1782L}, iterative sky-subtraction, coaddition using SWARP, and source detection using SEXTRACTOR \citep{2015MNRAS.449.2919L}. We calibrate against USNO-B1 and 2MASS.

RATIR observed the field of GRB 180620A from 2018-06-20 08:36:59.4 to 10:47:36.9 UTC (from $T + 0.03$ to $T + 2.21$ hours after the BAT trigger), obtaining a total of 1.78 hours exposure in the $r$ and $i$ bands \cite{22799}. RATIR obtains several set of frames with increasing exposure time, but for the first night we only used the 80~s exposures. For the second night, we combine all the frames for a total of 5.69 hours exposure. We also made some very deep stacks using late observations from July and August 2018 to gauge the brightness of the host galaxy.
Table~\ref{tab:datosratir} gives our RATIR photometry. For each image we list the initial time $t_i$, the final time $t_f$, the total exposure time $t_{exp}$ and, the \emph{r'}, \emph{i'} magnitudes (not corrected for Galactic extinction) with their the 1$\sigma$ total uncertainties (including both statistical and systematic contributions).

\subsection{Other Observations}
\label{sec:observationsterrestrial}

\cite{22812} reported photometry from the 0.76-m Katzman Automatic Imaging Telescope (KAIT) at Lick Observatory from $T+ 141$ ~s. They confirmed that the afterglow rises at early time and peaked around 300~s, as reported previously by \cite{22799} and \cite{22806}. GROND observed the field of GRB 180620A simultaneously in \emph{g'r'i'z'JHK}. Observations started at $T + 20$ hours. They report a reddening value $E_(B-V)$=0.11 in the direction of the burst \citep{22818}. Further optical photometry was published by \cite{22800,22803} and \citep{22811}.

\cite{22824} reported observations with Konus-Wind from which they produced a light curve shows a multi-peaked structure which starts at about $T+6.6$~s with total duration about 16~s (later confirmed by \citealt{22843}). The emission is seen up to about 2 MeV. The burst had a fluence of $9.80_{-1.02}^{+1.74}\times10^{-6}$ erg cm$^{-2}$ and a 64-ms peak flux, measured from $T + 7.504$~s, of  $2.69_{-0.82}^{+0.94}\times10^{-6}$ erg cm$^{-2}$ $s^{-1}$ (both in the 20 keV-10 MeV energy range).

\section{X-rays and Optical, Temporal and Spectral analysis}
\label{sec:analysis}

\subsection{Temporal Analysis}
\label{sec:temporalanalysis}

Figure~\ref{fig:lightcurve} shows the optical and X-rays light curves for GRB 180620A. The RATIR $r$ and $i$ filters have effective wavelengths of 618 and 760 nm respectively.
We can observe a very similar behaviour between the data for the RATIR $r$ and $i$ filters in Figures~\ref{fig:lightcurve}. 
The COATLI $w$ filter is well-described by $(w-r')=0.0256(g'-r')$ where the apostrophes in $g'$ and $r'$ refer to SDSS filters \citep{2019ApJ...872..118B}. Comparatively, the response of the COATLI $w$ filter is quite different to that of the Pan-STARRS1 $w$ filter \citep{2012ApJ...750...99T. The rising response from 400 to 650 nm is similar, but the Pan-STARRS1 response is then flat to the sharp filter cutoff at 800 nm, whereas the COATLI response falls from 700 nm out to the limit of the CCD response at 1100 nm.}

Therefore, we can consider this filter as an $r$ filter and fit the $RATIR-r'$ and $COATLI-w$ filters together. We increase the data sample by adding optical observations in $r$ from \cite{22811} and \citealt{22818}.

The prompt emission from the GRB detected by {\itshape Swift}/BAT lasted until about $T + 23$~s. The earliest complementary data are from COATLI at $T + 39.1$~s and RATIR at $T + 121.4$~s, and {\itshape Swift}/XRT starting at $T + 267$~s. Therefore, we focus our analysis on the early and late afterglow. 

We fit the light curves with power-laws segments $F_\nu \propto t^{-\alpha}$, in which $F_\nu$ is the flux density, $t$ is the time since the BAT trigger, and $\alpha$ is the temporal index. 
Once we fit the data, we compare our results with the theoretical description developed by
\cite{1998ApJ...497L..17S} and \cite{2002ApJ...568..820G}, who derived the synchrotron light curves when the outflow is decelerated by a constant-density interstellar medium (ISM) and the stellar wind of the progenitor, respectively. The power-law fits are summarized in Table~\ref{tab:fit}. 

We divide the light curve into three stages, labelled I, II, and III in Figure~\ref{fig:lightcurve}.

\begin{enumerate}

\item Stage I. This corresponds to the optical light curve for $t < 110$~s.

This region can be fitted with a temporal index of $\alpha_\mathrm{O,normal} = 0.86 \pm 0.12$. This  decay is similar to 
others seen
in many afterglows after the plateau phase (although in our case the plateau phase occurs at later times).
Unfortunately, there is no Swift/XRT data at early times to compare to the optical data. 

\item Stage II. This corresponds to $110 < t < 7800$~s. 

From $T+110$~s to $T+550$s we see a bright flare in optical with a peak at about $T + 270$~s. The duration of the flare $\Delta t$ over the peak time $t$ is $\Delta t/t=1.63$. Empirically, we fit two broken power-law segments in order to obtain parameters to describe the flare.  The rise has $\alpha_\mathrm{O,rise}=-0.83\pm0.07$ for $110 < t < 270$~s and the decay has $\alpha_\mathrm{O,decay}=1.32\pm0.01$ for $270 < t < 550$~s.   

The $X$-rays observations began at $T + 320$~s, after the peak of the flare in the optical. We do not see clear evidence for an X-ray counterpart of the flare, however data are noisy and we cannot completely exclude this possibility.

After the flare, from $T + 550$  to $T + 7800$~s, we see a plateau phase. The global fit of this stage can be described by a power-law with a temporal index in X-rays of $\alpha_\mathrm{X,plateau} = 0.40 \pm 0.03$ for $T<7800$~s whereas in optical with a temporal index $\alpha_\mathrm{O,plateau} = 0.41 \pm 0.02$ for $550<T<7800$~s. An extrapolation of the plateau to the end of Stage I suggests a 
transition from the decay in Stage I to the plateau in Stage II upon which the flare is superposed.

After the optical flare we observe similar fluctuations in both optical and $X$-rays. This suggests a common origin. The analysis and discussion of this stage are in \S~\ref{sec:fluctuations}.

\item Stage III. This corresponds to $T > 7800$~s. 

This region can be fitted as a power-law with a temporal index in X-rays of $\alpha_\mathrm{X,steep} = 2.17 \pm 0.13$ whereas in the optical the temporal index is $\alpha_\mathrm{O,steep} = 1.84 \pm 0.12$. These temporal indices are consistent with the post-jet break stage.
Nevertheless, the temporal evolution in the optical and X-rays differs only in 1.3 $\sigma$ and so it is not 
statistically significant.

\end{enumerate}

\subsection{Spectral Analysis}
\label{sec:spectrum}

\begin{figure}
\centering
 \includegraphics[width=0.5\textwidth]{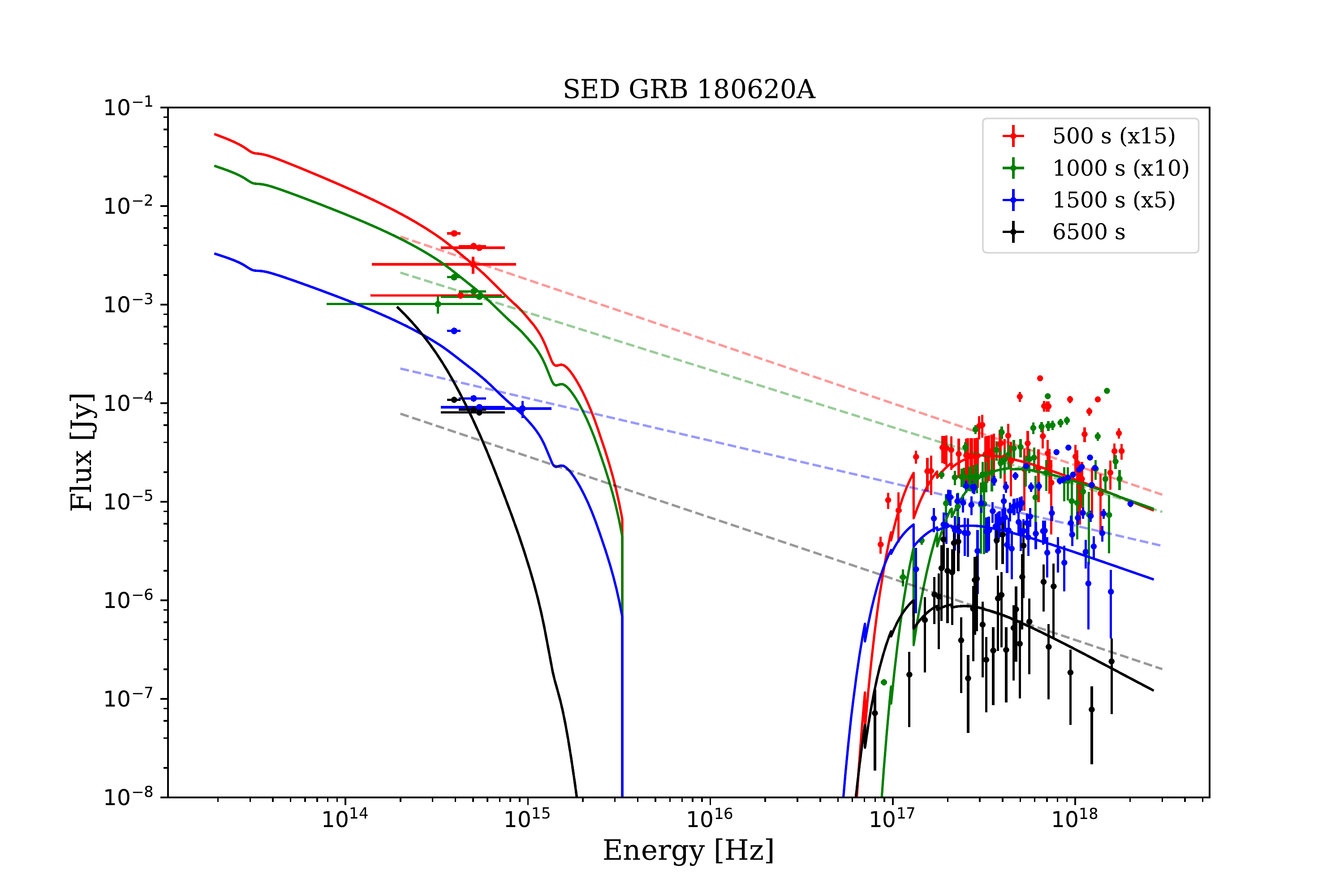}
 \caption{ The SED of GRB 180620A from X-rays to the optical. The data are from the COATLI telescope, the RATIR instrument and {\itshape Swift} instrument at $T=500$~s (red), $T=1000$~s (green), $T= 1500$~s (blue) and $T=6500$~s (black). We add the corresponding \emph{Xspec} model which takes into account the effects of redding and absorption by the dust (lines) and the linear fit (dotted line).}
 \label{fig:sed}
\end{figure}

We retrieved the {\itshape Swift/}XRT X-rays and {\itshape Swift/}UVOT optical spectrum at mean arrival times of $T= 500$~s, $T=1000$~s, $T= 1500$~s and $T= 6500$~s
from the online repository\footnote{http://www.swift.ac.uk/xrt\_spectra/. The exposure times are about 500~s for $T= 500$ s, $T= 1000$ s and $T= 1500$ s whereas for $T= 6500$~s is about 2000~s.}.
We add our COATLI and RATIR photometry at the corresponding mean photon arrival time. For X-rays data, we binned 2 individual values. Figure~\ref{fig:sed} shows the resulting combined spectral energy distributions for these epochs (SED).

We fit the photometry with a spectral power-law $F_\nu\propto \nu^{-\beta}$, in which $F_\nu$ is the flux density, $\nu$ is the frequency, and $\beta$ is the spectral index.
From the X-rays to the optical, the SED can be fitted with a simple power law in each case. In order to have a better constrain, we average the temporal indices obtained for each epoch and the resulting value is an  average index of $\beta=0.57\pm0.07$.  Under our assumption of a thin-shell evolving in the slow-cooling regime with the cooling break above the X-rays \citep{2007ApJ...655..391K,2016ApJ...818..190F, 2018ApJ...859...70F}, we would expect the spectral index to be $(p-1)/2$ or $0.57\pm0.08$ for $p=2.15\pm 0.16$ in a good agreement with the $\beta$ obtained; this value of $p$ is explained in more detail in the next section.

Figure~\ref{fig:sed} also shows the \emph{Xspec} model fit 
which takes into account the effects of reddening and absorption by the dust \citep{1996ASPC..101...17A}. We use a reddening of E(B-V) 0.13 \citep{22798},
the redshift limit calculated with the \emph{Swift/}UVOT of $z=1.2$ \citep{22805}, and the column density of $1.60\times10^{20}$cm$^{-2}$ \citep{22798}. We obtained a reduced $\chi^2<3.5$ in all the three different epochs reported. 

\section{Theoretical Interpretation }
\label{sec:interpretation}
The long-lived emission of afterglows is explained by a synchrotron forward-shock model. The distribution of accelerated electrons 
is assumed to be given by 
$N(\gamma_{\rm e}) \propto \gamma_{\rm e}^{-p}$ in which $\gamma_{\rm e}$ is the Lorentz factor of the electron and $p$ is the power-law index. The observed synchrotron flux is described by a series of power-law segments $F_{\nu}\propto t^{-\alpha}\nu^{-\beta}$ in time $t$ and frequency $\nu$. \cite{1998ApJ...497L..17S} and \cite{2000ApJ...543...66P} derived the synchrotron light curves when the outflow is decelerated by a constant-density interstellar medium (ISM) and the stellar wind of the progenitor. 

Most GRBs afterglows show three phases: a short plateau phase after the end of the prompt emission, followed by a ``normal'' decay phase, and finally a steeper decay after a jet break. The plateau phase has been explained by various models including jet coasting (i.e., the head of the relativistic jet moving at constant velocity) and  energy injection  \citep[see, for example, ][]{2006ApJ...642..389N,2006ApJ...642..354Z,2015ApJ...806..205D}. The normal decay is expected after the plateau \citep{1998ApJ...497L..17S,2000ApJ...545..807K} once the jet starts decelerating. Finally, the deceleration of the relativistic jet produces a jet break and subsequent steeper decay when $\Gamma\lesssim 1/\theta_{\rm jet}$, where $\Gamma$ and $\theta_{\rm jet}$ refer to the Lorentz factor and the jet opening angle respectively.

The order of the different light curve phases is unusual in GRB180620A.
At early times, we see the normal decay (Stage I) shown by most GRBs \citep{2006ApJ...642..354Z,2009MNRAS.397.1177E,2006ApJ...642..389N}, followed by a plateau (Stage II) in the optical and X-ray. At later times (Stage III), both light curves decrease steeply. 

We begin by analysing the initial normal decay phase in order to provide a constraint on the population of relativistic electrons. Subsequently, we will assume continuity in this population and consider the later phases and possible explanations for the optical and X-rays fluctuations.

\subsection{Initial Normal Decay}
\label{sec:normal}
For the optical, 
the temporal decay is consistent with
emission below the cooling frequency ($\nu_m<\nu<\nu_c$) in a slow-cooling scenario, and expect $F_\nu\propto \nu^{-(p-1)/2}t^{3(1-p)/4}$ for a jet being driven into a constant-density environment \citep{1998ApJ...497L..17S}. The observed value of $0.86 \pm 0.12$ gives $p=2.15\pm0.16$. This value of $p$ predicts a spectral index $\beta_\mathrm{O} = 0.57 \pm 0.08$, which is consistent with the observed value at later times derived in \S\ref{sec:spectrum}.

\subsection{Plateau phase}
GRB 180620A shows a quite long plateau in both X-rays and the optical. The long plateau starts no later than $T + 500$~s (its true start may be concealed by the flare) and ends with a common break in both X-rays and optical at $T+7800$~s. The simultaneous transition in both the X-rays and the optical to the subsequent late decay suggests the same origin for emission in both spectral regions.
 The time and spectral evolution of this phase ($F_\nu\propto  t^{-(0.47\pm0.1)}\,\nu^{-(0.58\pm0.08)}$) are consistent with energy injection as discussed in \S~\ref{sec:discussion}.
 
This phase can be understood within the standard external shock model \citep{1998ApJ...496L...1R,2006ApJ...642..354Z} as 
late-time energy injection
\citep[e.g.,][]{2019ApJ...872..118B}.

\subsection{Flare}

A striking optical flare occurs from $T+110$~s to $T+550$~s at the start of the plateau phase (stage II). One might naturally posit that both the plateau and the flare have the same physical origin in the central engine. However, such a neat link does not stand up to detailed scrutiny.

A flare caused by the central engine would be expected to have similar properties to the prompt emission \citep{2006ApJ...642..354Z}, with a steep rise and rapid decay and the presense of a bright X-ray flare. 
Our X-ray light curve starts towards the end of the flare, so we are not able to determine if a counterpart in X-rays is present. However, the steepness of the rise and fall in the optical give strong contraints.
We fit the rise with $\alpha_\mathrm{O,rise}=-0.83\pm0.07$ for $110 < T < 270$~s and the decay with $\alpha_\mathrm{O,decay}=1.32\pm0.01$ for $270 < T < 550$~s (Table~\ref{tab:fit}). These are too slow to be consistent with late central activity \citep{2006ApJ...642..354Z}, and so we rule this out as the origin of the optical flare in GRB 180620A.

Another possibility is that the flare is generated by fluctuations in the density of the environment. This idea was explored by  \cite{2002A&A...396L...5L}, who found that inhomogeneities could generate increases in flux by factors of up to about 3-4 compared to the uniform case, which matches the observed brightening in GRB 180620A of a factor of 2--3. Moreover, the time for flux peak $\Delta t$ over the peak time $t$, $\Delta t/t$ is about 2--4, and therefore supports the possibility that the flare present in GRB 180620A is produced as a result of a density bump. Nevertheless, the change in temporal indices $\Delta\alpha$ present in the flare  defined as $\Delta\alpha=\alpha_\mathrm{O,plateau}-\alpha_\mathrm{O,rise}=1.23\pm0.10$ indicates a sharp and large jump in a uniform density profile needed in order to produce the observable increase in the light curve of GRB 180620A. If the circumstellar density suddenly increases by a factor 100, it would produce a variation of $\Delta\alpha=1.0$ \citep{2014A&A...572A..55M} which is similar to the observed value. Physically, it is hard to conceive of a scenario in which the density increases so dramatically \citep{2007MNRAS.380.1744N}.

Therefore, we conclude that density fluctuations are not responsible for the flare observed in GRB 180620A.

Finally, we consider a reverse shock component. In this scenario, the synchrotron radiation from the reverse shock usually peaks in the optical  \citep{2019ApJ...881...12B} and shows slow temporal indices for rise and decay ($\alpha<-2$ and $\alpha<2$ respectively). This is consistent with our fitted indices for the rise and decay and with the possible absence of an X-ray counterpart.

In conclusion, we suggest a reverse shock component present in the light curve of GRB 180620A as the most likely explanation for the flare observed at $110 < t < 550$~s.

\subsection{Fluctuations}
\label{sec:fluctuations}

From $T+600$ to $T+10000$~s both the optical and X-ray light curves appear to show fluctuations above the smooth power-law fit. To evaluate their significance, we calculated the RMS deviation of the data about the fit, after first binning the data in groups of three observations to increase the signal-to-noise ratio. For the optical the RMS deviation is $41\sigma$, which indicates that the fluctuations are real and very strong. However, for X-rays
the larger noise of the data and lower amplitude of the deviations leads to a RMS deviation of only $1.4\sigma$, which makes it impossible to affirm or reject the idea of an X-rays counterpart for the strong optical fluctuations.

GRB 180620A is not the first event to show fluctuations like these. There are three scenarios to explain this kind of behaviour: late central activity, two-components jets, and density gradients. For example, \cite{2010MNRAS.406.2149M} analysed the GRB 100117A by assuming the first scenario. They interpreted the observed fluctuations in X-rays as flares produced by late central activity.
Another event which shows fluctuations is GRB 081028 which has three X-rays flares with different widths and maxima.
\citet{2010MNRAS.402...46M} suggested the presence of two distinct physical regions to explain the pulses observed in the light curve.
Finally, it is possible to consider a third scenario to explain these fluctuations: density gradients in the environment. These fluctuations are typically characterized by the presence of bright bumps on top of the usual power-law decay. This was the interpretation of \citet{2002A&A...396L...5L} for data of GRB 021004. Nevertheless, this effect might be negligible if the dominant component in this phase is the reverse shock.
	
Consistently with the plateau phase, we suggest that the optical multi-peaked fluctuations showed by GRB 180620A are produced also by late 
energy injection.

\subsection{Break}

We observed a break in the light curve around $T+7800$~s in both the optical and X-rays.

The first possibility to explain this break is the transition of the cooling frequency across the band \citep{1998ApJ...497L..17S} as a consequence of the end of a shallow decay or plateau phases in the X-rays and optical light curves. This scenario predicts different temporal indices of $\alpha_{O,steep}=3(1-p)/4$ and $\alpha_{X,steep}=(2-3p)/4$  for optical and X-rays after the break, respectively which corresponds to $\alpha_{O,steep}=0.86\pm 0.12$ and $\alpha_{X,steep}=1.11\pm 0.12$ using $p=2.15\pm0.16$.

Our observed values $\alpha_{O,steep}=1.84\pm 0.12$ and $\alpha_{X,steep}=2.17\pm 0.13$ respectively are much larger than the expected. 
Therefore, we conclude that we do not observe a transition of the cooling frequency.

Second, we consider the jet break, a geometrical and relativistic effect due of the bulk Lorentz factor $\Gamma$ decreasing with the time. The relativistic beaming present in GRBs, predicts that the jet is seen only within an angle of $1/\Gamma$ from the jet axis. Over time, the jet decelerates and $\Gamma$ decreases. We expect a steep temporal break when the $1/\Gamma$ angle becomes larger than the jet opening angle. We suggest that the break at the end of the plateau phase is a jet break. This is supported by the achromatic appearance of the break and also the subsequent decay. 

\subsection{Late Steep Decay}

For $t>7800$~s, we derive power-law segment temporal indices of $\alpha_\mathrm{O,steep}=1.84\pm0.12$ and $\alpha_\mathrm{X,steep}=2.17\pm0.13$ in the optical and X-rays respectively. We interpret this as the post jet-break decay phase. For this phase, both indices are in a good agreement as expected for an achromatic decay with $\alpha_{steep}$ of about $2$ \citep{2015PhR...561....1K}.

 \section{Discussion}
\label{sec:discussion}

GRB 180620A is definitively not a standard burst. Comparing it to the canonical GRB light curve, we see a very different temporal evolution. We see an initial normal decay phase, an optical flare starting at $T+110$~s and lasting 440 seconds (with a peak at 270~s), and a plateau phase from $T + 550$ to $T + 7800$~s with clear fluctuations. Finally, we observe the signature of the jet break at $T+7800$~s followed by a steeper achromatic decay.

We propose the following scenario to explain the optical and X-ray light curves. 
We interpret the normal decay observed during $40 \le t < 110$ s as produced by a decelerating forward shock component.
Seeing the normal decay at 40 seconds after the start of the burst implies a deceleration time $t_d = 200 \;(E_{53}/n)^{1/3} \Gamma_{0,2}^{-8/3} \lesssim 40$ s \citep{1998ApJ...497L..17S} (where $E_{53}$ is the equivalent kinetic energy in units of 10$^{53}$ ergs, $n$ is the circumstellar density and $\Gamma_0$ is the bulk Lorentz factor during the deceleration phase). This implies a circumstellar density $n\gtrsim 100$ cm$^{-3}$ (assuming an initial Lorentz factor $\Gamma_0\approx 100$ for the jet). 

During the deceleration phase, the Lorentz factor goes as $\Gamma\propto t^{-3/8}$ in the observer's frame. Thus, the bulk Lorentz factor decreases to $\Gamma = 100 \times (40 {\rm \; s}/t_d)^{-3/8}\approx 70 \; (t_d/40)^{3/8}$ at $t \approx 110$ s.

The flare observed from $T+110$~s to $T+550$~s and the following plateau phase can be then explained by assuming a late-time energy injection or reactivation due, e.g., by late central engine activity or by a Lorentz factor distribution of the ejecta. In this scenario, the flare is produced by the reverse shock component while the plateau is due to the reenergized forward shock emission. 
Finally, the late-time steeper decay is consistent with the post-jet break evolution which requires the energy injection to end, by coincidence, very close to the jet-break epoch, i.e. just before $T+7800$. A jet break at $T+7800$ implies a Lorentz factor $\Gamma_{\rm jb}\approx 1/\theta_{\rm jet} \approx 20$ for a jet opening angle $\theta_{\rm jet}\approx 3$~deg. Thus, the evolution of the jet Lorentz factor during the plateau phase is given by 
$\Gamma \approx 40 \;(t/110\; {\rm s})^{-0.3}$, which implies a shock energy that increases (in the observer's frame) as $E(t)\propto t^{\rm m}$ for $m=0.3$. This value is consistent with the evolution of the optical and X-ray emission during the plateau phase $F_\nu\propto t^{[-(3p-3)+m(p+3)]/4}\,\nu^{-(p-1)/2}= t^{-(0.47\pm0.1)}\,\nu^{-(0.58\pm0.08)}$ \citep{2006ApJ...642..354Z,1998ApJ...496L...1R,2015PhR...561....1K}.

 \section{Summary}
 \label{sec:summary}

We have presented optical photometry of the afterglow of GRB 180620A with the COATLI telescope and the RATIR instrument. COATLI received an automated alert at $T + 15.2$~s, and its quick response allowed us to obtain photometry of the early afterglow from $T + 39.3$~s, only 21 seconds after the alert. Furthermore, RATIR observed the field of GRB 180620A from 0.03 to 0.47 hours after the BAT trigger and one night after the trigger.

The unusual light curve of GRB 180620A can be explained by a model with late-time energy injection combined with a reverse shock component. Nevertheless, the unusual order of the phases in GRB 180620A are an excellent example of the diversity of GRB behaviour at early times.

Finally, we want to remark on the importance of observations by fast terrestrial telescope of the very early light curves of GRBS, prior even to pointed observations with Swift/XRT and Swift/UVOT, as they are indispensable for gaining a full understanding of the GRB phenomena.

\section*{acknowledgments}

We thank Enrique Moreno and Diego L\'opez C\'amara for useful discussions. We thank the staff of the Observatorio Astron\'omico Nacional on Sierra San Pedro M\'artir. Some of the data used in this paper were acquired with the COATLI telescope and interim instrument at the Observatorio Astron\'omico Nacional on the Sierra de San Pedro M\'artir. COATLI is funded by CONACyT (LN 232649, 260369, and 271117) and the Universidad Nacional Aut\'onoma de M\'exico (CIC and DGAPA/PAPIIT IT102715, IG100414, IN109408, and IN109418) and is operated and maintained by the Observatorio Astron\'omico Nacional and the Instituto de Astronom\'ia of the Universidad Nacional Aut\'onoma de M\'exico.
RATIR is a collaboration between the University of California, the Universidad Nacional Auton\'oma de M\'exico, NASA Goddard Space Flight Center, and Arizona State University, benefiting from the loan of an H2RG detector and hardware and software support from Teledyne Scientific and Imaging. RATIR, the automation of the Harold L. Johnson Telescope of the Observatorio Astron\'omico Nacional on Sierra San Pedro M\'artir, and the operation of both are funded through NASA grants NNX09AH71G, NNX09AT02G, NNX10AI27G, and NNX12AE66G, CONACyT grants INFR-2009-01-122785 and CB-2008-101958, UNAM PAPIIT grants IG100414 and IA102917, UC MEXUS-CONACyT grant CN 09-283, and the Instituto de Astronom{\'\i}a of the Universidad Nacional Auton\'oma de M\'exico. We acknowledge the vital contributions of Neil Gehrels and Leonid Georgiev to the early development of RATIR.
FDC aknowledges support from the UNAM-PAPIIT grant IN117917

This work made use of data supplied by the UK Swift Science Data Centre at the University of Leicester.

\startlongtable
\begin{deluxetable}{rrrcc}
\tablecaption{COATLI observations of GRB 180620A\label{tab:datoscoatli}}
\tablehead{\colhead{$t_i$ (s)}& \colhead{$t_f$ (s)}& \colhead{$t_{exp}$ (s)}&\colhead{w}}
\startdata
39.3 & 44.0 & 5 & 16.94 $\pm$ 0.06 \\
48.3 & 53.0 & 5 & 17.23 $\pm$ 0.07 \\
57.3 & 62.0 & 5 & 17.50 $\pm$ 0.09 \\
66.3 & 71.0 & 5 & 17.47 $\pm$ 0.09 \\
75.3 & 80.0 & 5 & 17.53 $\pm$ 0.09 \\
84.3 & 89.0 & 5 & 17.68 $\pm$ 0.10 \\
93.3 & 98.0 & 5 & 17.66 $\pm$ 0.10 \\
102.3 & 107.0 & 5 & 17.74 $\pm$ 0.11 \\
111.3 & 116.0 & 5 & 17.66 $\pm$ 0.10 \\
120.3 & 125.0 & 5 & 17.68 $\pm$ 0.10 \\
141.2 & 146.0 & 5 & 17.39 $\pm$ 0.08 \\
150.2 & 155.0 & 5 & 17.28 $\pm$ 0.07 \\
159.2 & 164.0 & 5 & 17.53 $\pm$ 0.09 \\
169.0 & 174.0 & 5 & 17.44 $\pm$ 0.09 \\
178.1 & 183.0 & 5 & 17.30 $\pm$ 0.07 \\
187.1 & 192.0 & 5 & 17.22 $\pm$ 0.07 \\
196.1 & 201.0 & 5 & 17.33 $\pm$ 0.07 \\
205.1 & 210.0 & 5 & 17.14 $\pm$ 0.07 \\
214.1 & 219.0 & 5 & 17.14 $\pm$ 0.07 \\
223.1 & 228.0 & 5 & 17.07 $\pm$ 0.06 \\
243.9 & 249.0 & 5 & 16.99 $\pm$ 0.06 \\
252.9 & 258.0 & 5 & 16.92 $\pm$ 0.05 \\
261.9 & 267.0 & 5 & 16.95 $\pm$ 0.06 \\
270.9 & 276.0 & 5 & 16.94 $\pm$ 0.06 \\
279.9 & 285.0 & 5 & 16.96 $\pm$ 0.06 \\
289.0 & 294.0 & 5 & 17.09 $\pm$ 0.06 \\
298.0 & 303.0 & 5 & 17.06 $\pm$ 0.06 \\
307.0 & 312.0 & 5 & 17.04 $\pm$ 0.06 \\
316.0 & 321.0 & 5 & 17.10 $\pm$ 0.06 \\
325.0 & 330.0 & 5 & 17.09 $\pm$ 0.06 \\
346.9 & 352.0 & 5 & 17.26 $\pm$ 0.07 \\
355.9 & 361.0 & 5 & 17.30 $\pm$ 0.08 \\
364.9 & 370.0 & 5 & 17.36 $\pm$ 0.08 \\
373.9 & 379.0 & 5 & 17.43 $\pm$ 0.08 \\
382.9 & 388.0 & 5 & 17.38 $\pm$ 0.08 \\
391.9 & 397.0 & 5 & 17.45 $\pm$ 0.09 \\
400.9 & 406.0 & 5 & 17.42 $\pm$ 0.08 \\
410.0 & 415.0 & 5 & 17.36 $\pm$ 0.08 \\
419.0 & 424.0 & 5 & 17.46 $\pm$ 0.08 \\
428.0 & 433.0 & 5 & 17.64 $\pm$ 0.10 \\
450.3 & 455.0 & 5 & 17.87 $\pm$ 0.12 \\
459.3 & 464.0 & 5 & 17.67 $\pm$ 0.10 \\
468.3 & 473.0 & 5 & 17.51 $\pm$ 0.09 \\
477.3 & 482.0 & 5 & 17.82 $\pm$ 0.12 \\
486.3 & 491.0 & 5 & 17.88 $\pm$ 0.12 \\
495.4 & 500.0 & 5 & 17.80 $\pm$ 0.11 \\
504.4 & 509.0 & 5 & 17.81 $\pm$ 0.11 \\
513.4 & 518.0 & 5 & 17.93 $\pm$ 0.13 \\
522.4 & 527.0 & 5 & 18.03 $\pm$ 0.14 \\
531.4 & 536.0 & 5 & 17.88 $\pm$ 0.12 \\
553.0 & 634.0 & 81 & 17.71 $\pm$ 0.00 \\
759.0 & 841.0 & 245 & 18.37 $\pm$ 0.01 \\
1104.7 & 1187.0 & 246 & 18.61 $\pm$ 0.01 \\
1486.3 & 1568.0 & 245 & 18.50 $\pm$ 0.01 \\
1795.0 & 1877.0 & 246 & 18.54 $\pm$ 0.01 \\
2101.3 & 2183.0 & 244 & 18.53 $\pm$ 0.01 \\
2679.0 & 3033.0 & 354 & 18.65 $\pm$ 0.01 \\
3079.0 & 3436.0 & 357 & 18.71 $\pm$ 0.02 \\
3482.0 & 3835.0 & 353 & 18.80 $\pm$ 0.01 \\
3881.0 & 4239.0 & 358 & 18.81 $\pm$ 0.01 \\
4285.0 & 4638.0 & 353 & 18.84 $\pm$ 0.01 \\
4685.0 & 5042.0 & 357 & 18.89 $\pm$ 0.01 \\
5087.0 & 5443.0 & 356 & 18.98 $\pm$ 0.01 \\
5489.0 & 5846.0 & 357 & 18.98 $\pm$ 0.02 \\
5891.0 & 6246.0 & 355 & 19.01 $\pm$ 0.01 \\
6552.0 & 6906.0 & 354 & 19.10 $\pm$ 0.01 \\
6952.0 & 7308.0 & 356 & 19.07 $\pm$ 0.01 \\
7354.0 & 7708.0 & 354 & 19.05 $\pm$ 0.02 \\
7754.0 & 8111.0 & 357 & 19.27 $\pm$ 0.02 \\
8157.0 & 8512.0 & 355 & 19.27 $\pm$ 0.03 \\
8961.0 & 9317.0 & 1067 & 19.41 $\pm$ 0.01 \\

\enddata
\end{deluxetable}

\startlongtable
\begin{deluxetable}{rrrcc}
\tablecaption{RATIR observations of GRB 180620A\label{tab:datosratir}}
\tablehead{\colhead{$t_i$ (s)}& \colhead{$t_f$ (s)}& \colhead{$t_{exp}$ (s)}&\colhead{r'}&\colhead{i'}}
\startdata
121.4 & 201.4 & 80 & 17.51 $\pm$ 0.01 & 17.21 $\pm$ 0.01 \\
214.1 & 294.1 & 80 & 17.21 $\pm$ 0.01 & 16.89 $\pm$ 0.01 \\
302.4 & 382.4 & 80 & 17.32 $\pm$ 0.01 & 17.00 $\pm$ 0.01 \\
394.4 & 474.4 & 80 & 17.56 $\pm$ 0.01 & 17.28 $\pm$ 0.01 \\
500.7 & 580.7 & 80 & 17.84 $\pm$ 0.01 & 17.53 $\pm$ 0.01 \\
606.4 & 686.4 & 80 & 18.12 $\pm$ 0.01 & 17.77 $\pm$ 0.01 \\
704.6 & 784.6 & 80 & 18.28 $\pm$ 0.02 & 17.98 $\pm$ 0.01 \\
826.2 & 906.2 & 80 & 18.34 $\pm$ 0.02 & 18.05 $\pm$ 0.01 \\
921.8 & 1001.8 & 80 & 18.43 $\pm$ 0.02 & 18.12 $\pm$ 0.02 \\
1013.9 & 1093.9 & 80 & 18.55 $\pm$ 0.02 & 18.20 $\pm$ 0.02 \\
1112.4 & 1192.4 & 80 & 18.59 $\pm$ 0.02 & 18.30 $\pm$ 0.02 \\
1228.9 & 1308.9 & 80 & 18.63 $\pm$ 0.02 & 18.31 $\pm$ 0.02 \\
1316.5 & 1396.5 & 80 & 18.60 $\pm$ 0.02 & 18.31 $\pm$ 0.02 \\
1411.5 & 1491.5 & 80 & 18.66 $\pm$ 0.02 & 18.31 $\pm$ 0.02 \\
1500.0 & 1580.0 & 80 & 18.65 $\pm$ 0.02 & 18.32 $\pm$ 0.02 \\
1599.0 & 1679.0 & 80 & 18.56 $\pm$ 0.02 & \nodata \\
1599.9 & 1679.9 & 80 & \nodata & 18.27 $\pm$ 0.02 \\
1712.5 & 1792.5 & 80 & 18.50 $\pm$ 0.02 & 18.23 $\pm$ 0.02 \\
1804.4 & 1884.5 & 80 & 18.56 $\pm$ 0.02 & \nodata \\
1804.5 & 1884.5 & 80 & \nodata & 18.26 $\pm$ 0.02 \\
1910.1 & 1990.1 & 80 & 18.58 $\pm$ 0.02 & 18.23 $\pm$ 0.02 \\
2010.9 & 2090.9 & 80 & 18.46 $\pm$ 0.02 & 18.24 $\pm$ 0.02 \\
2098.9 & 2178.9 & 80 & 18.45 $\pm$ 0.02 & \nodata \\
2099.0 & 2178.9 & 80 & \nodata & 18.11 $\pm$ 0.02 \\
2209.5 & 2289.5 & 80 & 18.56 $\pm$ 0.02 & 18.29 $\pm$ 0.02 \\
2305.6 & 2385.6 & 80 & 18.62 $\pm$ 0.02 & \nodata \\
2305.8 & 2385.8 & 80 & \nodata & 18.30 $\pm$ 0.02 \\
2403.7 & 2483.7 & 80 & 18.62 $\pm$ 0.02 & 18.30 $\pm$ 0.02 \\
2491.1 & 2571.1 & 80 & 18.58 $\pm$ 0.02 & 18.34 $\pm$ 0.02 \\
2583.9 & 2663.9 & 80 & \nodata & 18.36 $\pm$ 0.02 \\
2585.4 & 2665.4 & 80 & 18.68 $\pm$ 0.02 & \nodata \\
2681.6 & 2761.6 & 80 & 18.69 $\pm$ 0.02 & 18.37 $\pm$ 0.02 \\
2780.7 & 2860.7 & 80 & 18.65 $\pm$ 0.02 & 18.41 $\pm$ 0.02 \\
2868.1 & 2948.1 & 80 & 18.70 $\pm$ 0.02 & 18.41 $\pm$ 0.02 \\
2959.7 & 3039.7 & 80 & 18.70 $\pm$ 0.02 & 18.49 $\pm$ 0.02 \\
3047.4 & 3127.4 & 80 & 18.76 $\pm$ 0.02 & 18.47 $\pm$ 0.02 \\
3163.9 & 3243.9 & 80 & 18.71 $\pm$ 0.02 & 18.47 $\pm$ 0.02 \\
3275.7 & 3355.8 & 80 & 18.80 $\pm$ 0.02 & 18.50 $\pm$ 0.02 \\
3377.1 & 3457.1 & 80 & 18.83 $\pm$ 0.02 & 18.52 $\pm$ 0.02 \\
3466.3 & 3546.2 & 80 & 18.83 $\pm$ 0.02 & 18.58 $\pm$ 0.02 \\
3591.3 & 3671.3 & 80 & 18.83 $\pm$ 0.02 & 18.51 $\pm$ 0.02 \\
3679.7 & 3759.7 & 80 & 18.82 $\pm$ 0.02 & 18.54 $\pm$ 0.02 \\
3771.7 & 3851.7 & 80 & 18.90 $\pm$ 0.03 & 18.57 $\pm$ 0.02 \\
3868.9 & 3948.9 & 80 & 18.91 $\pm$ 0.02 & \nodata \\
3869.2 & 3949.2 & 80 & \nodata & 18.59 $\pm$ 0.02 \\
3968.9 & 4048.9 & 80 & 18.82 $\pm$ 0.02 & 18.59 $\pm$ 0.02 \\
4056.4 & 4136.4 & 80 & 18.86 $\pm$ 0.02 & 18.59 $\pm$ 0.02 \\
4148.9 & 4228.9 & 80 & 18.90 $\pm$ 0.02 & 18.61 $\pm$ 0.02 \\
4246.0 & 4326.0 & 80 & 18.94 $\pm$ 0.03 & 18.62 $\pm$ 0.02 \\
4338.3 & 4418.3 & 80 & 18.87 $\pm$ 0.03 & \nodata \\
4339.2 & 4419.2 & 80 & \nodata & 18.63 $\pm$ 0.02 \\
4428.9 & 4508.9 & 80 & 18.87 $\pm$ 0.03 & 18.64 $\pm$ 0.02 \\
4539.1 & 4619.1 & 80 & 18.95 $\pm$ 0.02 & 18.64 $\pm$ 0.02 \\
4626.5 & 4706.6 & 80 & 18.96 $\pm$ 0.03 & \nodata \\
4626.6 & 4706.6 & 80 & \nodata & 18.67 $\pm$ 0.02 \\
4727.3 & 4807.3 & 80 & \nodata & 18.66 $\pm$ 0.02 \\
4728.2 & 4808.2 & 80 & 18.96 $\pm$ 0.03 & \nodata \\
4835.7 & 4915.7 & 80 & 18.93 $\pm$ 0.03 & 18.66 $\pm$ 0.02 \\
4927.6 & 5007.6 & 80 & 19.02 $\pm$ 0.03 & 18.64 $\pm$ 0.02 \\
5035.5 & 5115.5 & 80 & 19.05 $\pm$ 0.03 & 18.68 $\pm$ 0.02 \\
5152.8 & 5232.8 & 80 & 18.93 $\pm$ 0.03 & 18.67 $\pm$ 0.02 \\
5251.3 & 5331.3 & 80 & 18.94 $\pm$ 0.03 & 18.69 $\pm$ 0.02 \\
5343.5 & 5423.5 & 80 & 18.97 $\pm$ 0.03 & 18.73 $\pm$ 0.02 \\
5439.7 & 5519.7 & 80 & 18.99 $\pm$ 0.03 & 18.73 $\pm$ 0.02 \\
5539.6 & 5619.6 & 80 & 18.92 $\pm$ 0.03 & 18.68 $\pm$ 0.02 \\
5627.0 & 5707.0 & 80 & 18.96 $\pm$ 0.03 & 18.67 $\pm$ 0.02 \\
5737.4 & 5817.4 & 80 & 18.99 $\pm$ 0.02 & 18.71 $\pm$ 0.02 \\
5833.8 & 5913.9 & 80 & 19.00 $\pm$ 0.02 & \nodata \\
5833.9 & 5913.9 & 80 & \nodata & 18.76 $\pm$ 0.02 \\
5949.2 & 6029.2 & 80 & 18.99 $\pm$ 0.03 & 18.76 $\pm$ 0.03 \\
6036.7 & 6116.7 & 80 & 18.96 $\pm$ 0.03 & \nodata \\
6036.8 & 6116.8 & 80 & \nodata & 18.76 $\pm$ 0.03 \\
6128.6 & 6208.6 & 80 & 19.02 $\pm$ 0.03 & 18.73 $\pm$ 0.02 \\
6216.1 & 6296.1 & 80 & 19.04 $\pm$ 0.03 & 18.79 $\pm$ 0.02 \\
6314.9 & 6394.9 & 80 & 19.05 $\pm$ 0.03 & 18.81 $\pm$ 0.02 \\
6426.0 & 6506.0 & 80 & 19.00 $\pm$ 0.03 & 18.74 $\pm$ 0.03 \\
6536.0 & 6616.0 & 80 & 19.05 $\pm$ 0.03 & 18.81 $\pm$ 0.02 \\
6624.7 & 6704.7 & 80 & 19.08 $\pm$ 0.03 & 18.81 $\pm$ 0.02 \\
6723.8 & 6803.8 & 80 & 19.06 $\pm$ 0.03 & 18.80 $\pm$ 0.02 \\
6812.3 & 6892.3 & 80 & 19.08 $\pm$ 0.03 & 18.79 $\pm$ 0.02 \\
6904.3 & 6984.3 & 80 & 19.14 $\pm$ 0.03 & 18.81 $\pm$ 0.02 \\
7000.3 & 7080.3 & 80 & 19.16 $\pm$ 0.03 & 18.83 $\pm$ 0.02 \\
7098.6 & 7178.6 & 80 & 19.10 $\pm$ 0.03 & 18.83 $\pm$ 0.02 \\
7184.8 & 7264.8 & 80 & 19.09 $\pm$ 0.03 & 18.88 $\pm$ 0.03 \\
7277.9 & 7357.9 & 80 & 19.10 $\pm$ 0.03 & \nodata \\
7278.0 & 7358.0 & 80 & \nodata & 18.91 $\pm$ 0.03 \\
7374.9 & 7454.9 & 80 & 19.18 $\pm$ 0.03 & 18.91 $\pm$ 0.03 \\
7492.0 & 7572.0 & 80 & 19.16 $\pm$ 0.03 & \nodata \\
7492.9 & 7572.9 & 80 & \nodata & 18.91 $\pm$ 0.03 \\
7580.5 & 7660.5 & 80 & 19.13 $\pm$ 0.03 & 18.88 $\pm$ 0.03 \\
7673.0 & 7753.0 & 80 & 19.16 $\pm$ 0.03 & 18.93 $\pm$ 0.03 \\
7779.5 & 7859.5 & 80 & 19.19 $\pm$ 0.03 & 18.90 $\pm$ 0.03 \\
7878.9 & 7958.9 & 80 & 19.19 $\pm$ 0.03 & 18.83 $\pm$ 0.02 \\
69480.0 & 95220.0 & 20484 & 23.58 $\pm$ 0.12 & 23.56 $\pm$ 0.11 \\

\enddata
\end{deluxetable}

\startlongtable
\begin{deluxetable*}{ccccr}

\tablecaption{Temporal and Spectral Power-Law Indices\label{tab:fit}}
\tablehead{\colhead{Region}&\colhead{Stage}&\colhead{Time Interval}&\colhead{Parameter}&{Value}}
\startdata
I&Normal Decay& $T<110$&$\alpha_\mathrm{O,normal}$&$0.86\pm0.12$\\[2ex]
  \hline
 & & $110<T<t_p$&$\alpha_\mathrm{O,rise}$&$-0.83\pm 0.07$\\
 &Optical-flare &270&$t_p$&\\
 II& & $t_p<T<550$&$\alpha_\mathrm{O,decay}$&$1.32\pm 0.01$\\[2ex]
&Plateau& 320$<T<$7800&$\alpha_\mathrm{X,plateau}$&$0.40\pm0.03$\\
&& 480$<T<$7800&$\alpha_\mathrm{O}$&$0.41\pm0.02$\\
 &decay&&$\beta_\mathrm{O}$&$0.56\pm0.08$\\[2ex]
 \hline\\
&Break& 7800&$t_j$&\\[2ex]
  \hline\\
 III&Late Steep&t$>7800$ &$\alpha_\mathrm{O,steep}$&$1.84\pm0.12$\\
&decay&&$\alpha_\mathrm{X,steep}$&$2.17\pm0.13$\\
 \enddata
\end{deluxetable*}

\end{document}